# Growth of (111)-textured $SrTiO_3$ thin films on Pt(111)/$Al_2O_3$(1-102) substrates by rf magnetron sputter deposition

Ekaterina Taylor[a,c], Ethan R. Cronk[b,c], Kalani Perera[b,c], Nuri W. Emanetoglu[a,c], Nicholas S. Bingham[b,c]

a. Department of Electrical and Computer Engineering, University of Maine
b. Department of Physics and Astronomy, University of Maine
c. Frontier Institute for Research in Sensor Technologies, University of Maine

## Abstract

$SrTiO_3$ thin films (20-30nm) with high-quality crystal structure and low roughness can be used as growth templates for complex oxides or as the dielectric materials for capacitor structures. In this work, stoichiometric STO thin films were grown by radio frequency (RF) magnetron sputtering on Pt(111) templates grown on r-$Al_2O_3$ substrates, and the effects of growth parameters as well as the underlying Pt template's structural properties on the quality of STO films were studied. A comparison between room temp grown Pt and Pt grown at elevated temps showed that the latter lead to highly ordered Pt and STO films with quasi 2D surface roughness. X-ray diffraction showed that the STO thin films had (111) and (222) crystal orientation. A rocking curve full width at half maximum value of 0.2° was achieved, indicating that the STO films are high-quality. The STO films have a roughness less than 1nm, as measured using atomic force microscopy, comparable to the underlying Pt template's roughness.

## Introduction

$SrTiO_3$ (STO) thin films have attracted significant attention due to their high dielectric constant, exceeding 300 at room temperature for bulk STO[1], and their stable dielectric constant at high frequencies (10-1000 MHz)[2]. STO has a paraelectric phase with a cubic perovskite structure at room temperature, while the ferroelectric phase emerges below a $T_C$ of 23K[3], resulting in negligible polarization hysteresis under standard device operating conditions and enabling potential operation of STO-based capacitors at elevated temperatures (up to ~700 °C), depending on the underlying electrode and template materials. Pt is a preferred bottom electrode material for high temperature applications, due to its lattice constant (3.92 Å) being very close to that of STO (3.905 Å).

STO films are typically deposited on Pt electrodes using physical and chemical methods, including magnetron sputtering [4-7], pulsed laser deposition (PLD)[8], atomic layer deposition (ALD)[9], and chemical vapor deposition (CVD)[10]. The films generally maintain a cubic perovskite structure with lattice parameters of ~3.90–3.905 Å, consistent with bulk STO. X-ray diffraction (XRD) studies reveal characteristic (100), (110), or (111) reflections depending on substrate orientation and deposition conditions.

The surface morphology measured by atomic force microscopy (AFM) in such studies typically exhibits root-mean-square (RMS) roughness values ranging from 4 nm up to 8 nm for textured STO films grown directly on Pt/$Al_2O_3$, with roughness increasing as grain faceting develops at higher growth temperatures[11-12]. It has been reported that STO/Ru and STO/Ru/$GeO_2$ heterostructures demonstrate a reduction in pinhole density and suppression of amorphous STO formation upon incorporation of $GeO_2$, with corresponding root-mean-square (RMS) surface roughness values being 1.6 nm and 0.8 nm, respectively[13]. These values contrast with atomically smooth surfaces (RMS $< 0.2$ nm) that can be achieved for single-crystal STO substrates[14]. Due to the close lattice match and similar symmetry between Pt(111) and STO(111), diffraction peaks can appear overlapping or closely spaced, often requiring high-resolution XRD or pole-figure analysis to clearly distinguish the STO film from the underlying Pt electrode[13].

These results indicate that, although $SrTiO_3$ (STO) can serve as a template for subsequent material growth or as a dielectric layer in capacitor structures, surface roughness and artifacts associated with deposition on Pt substrates currently limit its reliable application in films with thicknesses below 50 nm, indicating that further optimization is required. In this context, the present study examines the dependence of STO structural quality on the characteristics of the underlying Pt layer. When the Pt surface exhibits sufficiently high crystalline quality and smoothness, the growth of STO films with near-epitaxial characteristics and atomically smooth surfaces have been achieved for thicknesses of below 20-30 nm.

## Experimental Details

STO/Pt structures were grown on r-plane sapphire (r-$Al_2O_3$) substrates. The surface treatment of the substrates consisted of wet chemical cleaning (acetone, isopropanol, ethanol) and plasma pre-treatment for $Al_2O_3$ before Pt growth. The Pt growth template, which also served as the bottom electrode, and the following STO thin films were then grown using two different sputtering systems. Pt thin films (100nm thick) were deposited by AJA International ATC Orion Series deposition system on the r-$Al_2O_3$ substrates at 600°C with a plasma power of either 90 W or 180 W. STO films were then grown using RF sputtering in a custom-built chamber with a Advanced Energy Industries RF power supply, a Tempco Temperature controller, AJA International substrate heater and sputter guns, and an MKS Instruments Four-Channel Gas Controller.

*Table 1: Sputtering parameters*

| **Power [W]** | **Distance to substrate** | **Thickness(nm)** | **Growth Rate** |
|---|---|---|---|
| 20 | 14 | 30-40 | 1-1.3 Å/min |
| 30 | 14 | 35-45 | 1.1-1.5 Å/min |
| 40 | 14 | 50-60 | 1.6-2 Å/min |
| 20 | 12 | 20-30 | 0.67-0.8 Å/min |
| 30 | 12 | 30-40 | 1-1.3 Å/min |
| 40 | 12 | 45-55 | 1.5-1.8 Å/min |

The substrate growth temperature was 300°C, with a pressure of 3 mTorr of Ar and 7 mTorr of $O_2$, using a stoichiometric STO target (Princeton Scientific) 12 cm and 14 cm away from the substrates. The power settings were either 20 W, 30 W or 40 W. Growth time was kept constant at 5 hours. After growth, the samples were in-situ annealed in 300+ mTorr of $O_2$ at 550°C for 1 hour to improve the crystallinity of the STO thin films. The growth conditions and resulting growth rates are summarized in Table 1.

The stoichiometry of STO films was investigated using X-ray photoelectron spectroscopy (XPS, SPECS Phoibos hemispherical analyser under $10^{-9}$ Torr vacuum), comparing the peaks for Sr, Ti and O. Microstructural characterization methods included X-ray diffraction (XRD, Malvern Panalytical X'Pert 3 Pro MRD diffractometer with 1.54 Å Cu Ka radiation) θ-2θ scans, rocking curve XRD and RHEED. AFM was used to investigate the surface structure of the Pt and STO/Pt film.

## Results and Discussion

*Composition Analysis*

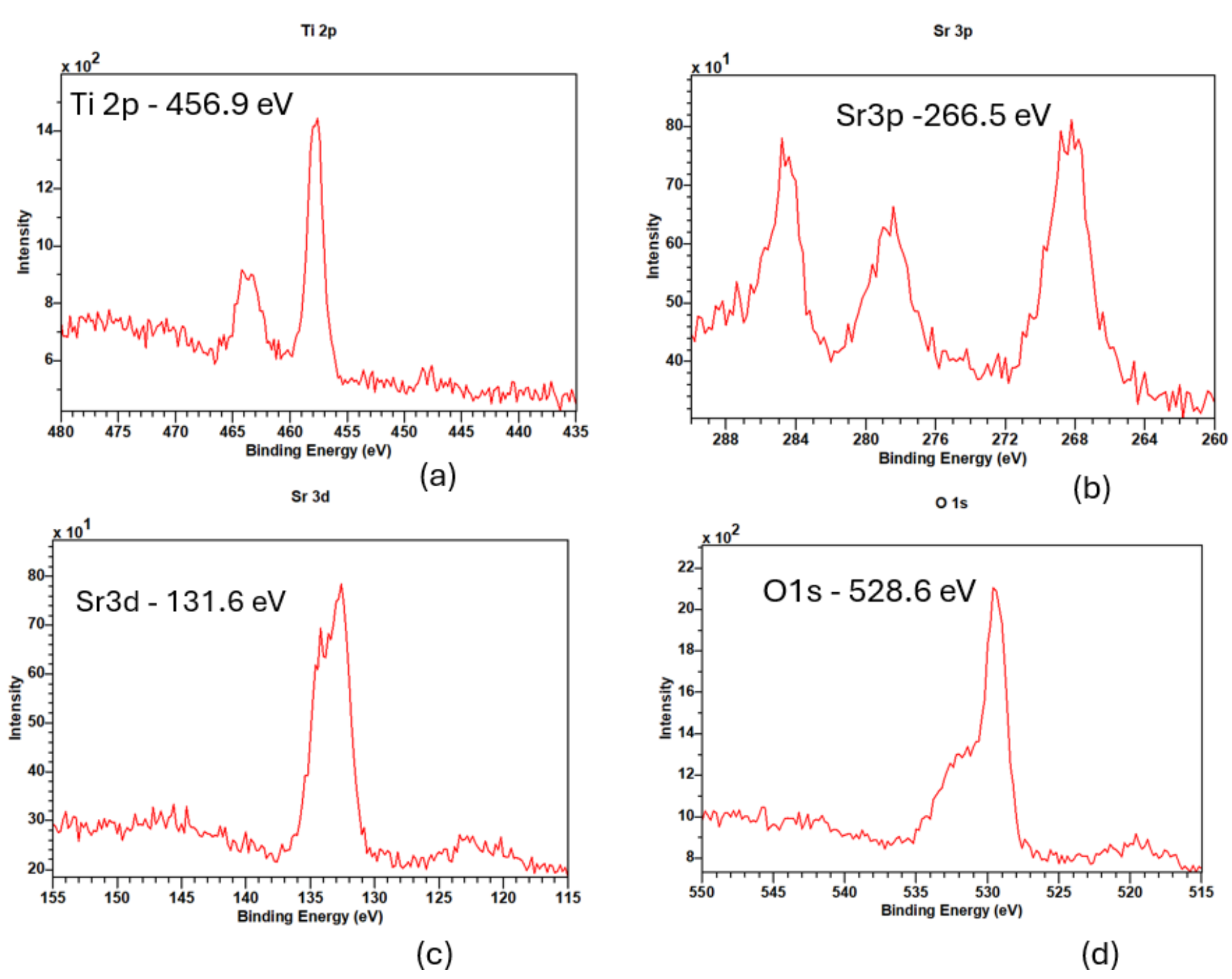

*Figure 1. X-ray photoelectron spectroscopy for 20W 14cm STO film: (a) Sr 3d peak (b) Sr 3p peak (c) Ti 2p peak (d) O 1s peak*

XPS was used to study the chemical composition of the STO films. The XPS survey spectrum presented in Figure 1 shows pronounced photoelectron peaks corresponding to Sr, Ti, and O. A Pt signal originating from the underlying template is also detected from the uncovered side of sample.

Additional signals from C, Al, and Na are attributed to the side of sapphire and sample holder. The hydrocarbon C 1s signal at 284.9 eV is used as the energy reference to correct for charging. Table 2 summarizes the binding energies of the main elements; the Ti 2p peak is observed at approximately 459 eV. This binding energy is consistent with reported values for Ti in $SrTiO_3$ thin films (≈458–459 eV). Variations in the surface atomic concentrations of Sr, Ti, and O are observed, and fall in agreement with previously reported XPS studies[15].

*Table 2: Spectral features for main elements of STO*

| **Element/ Transition** | **Peak energy (eV)** | **Peak Area (counts)** | **FWHM (eV)** | **Sensitivity** | **Concentration ±10%error** |
|---|---|---|---|---|---|
| Sr 3p | 266.5 | 0.0019 | 2.627 | 5.488 | 25.7 |
| Ti 2p | 456.9 | 0.0015 | 1.475 | 4.690 | 23.7 |
| O 1s | 528.6 | 0.0017 | 1.84 | 2.494 | 50.6 |
| C 1s | 284.9 | 0.0006 | 3.75 | 1.000 | |

*Film orientation and crystal structure quality*

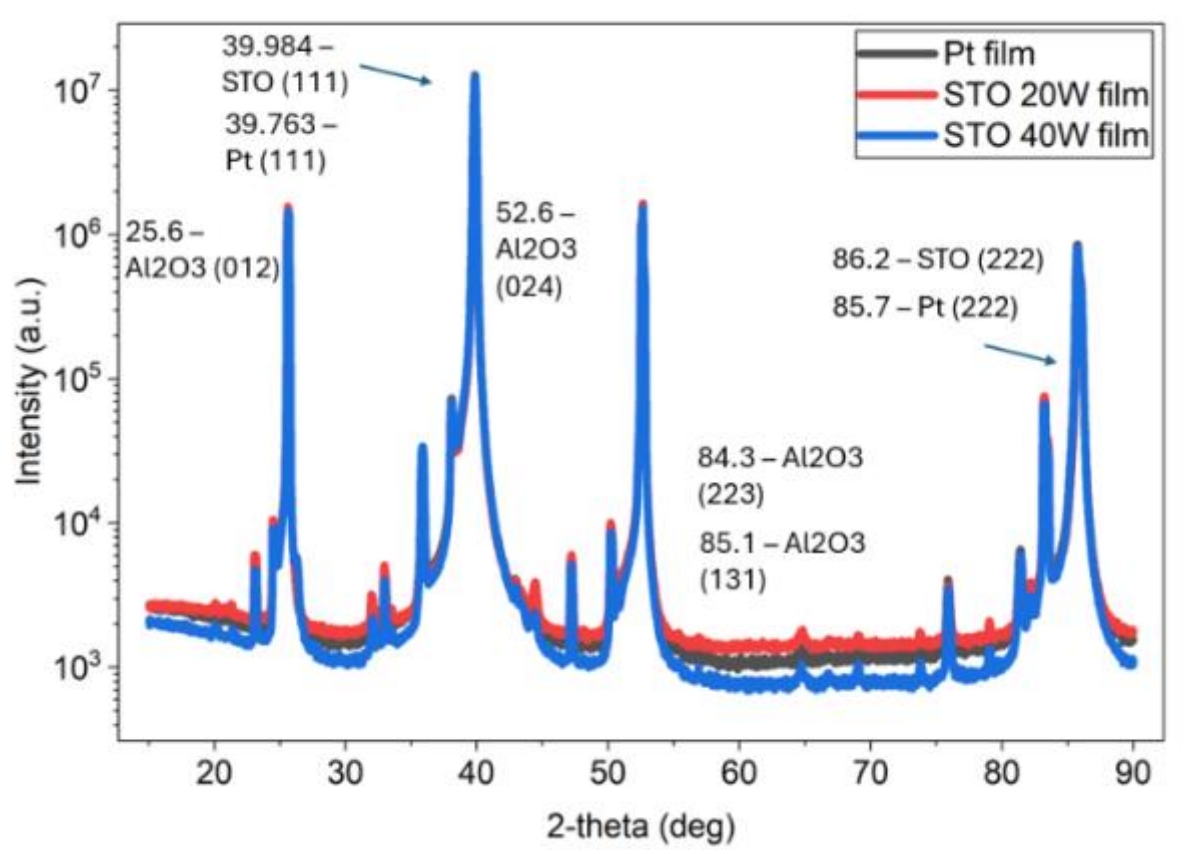

*Figure 2 Characteristics of $SrTiO_3$ and Pt thin films deposited on Pt/Al2O3 by RF magnetron sputtering at various power.*

Figure 2 compares the XRD patterns of 30 nm and 60 nm thick STO films grown on 100 nm Pt layers, using 20 W and 40 W plasma power respectively, with that of a 100 nm Pt reference sample without STO. The Pt layer exhibits a strong (111) preferred orientation, while the STO films show predominantly (111) out-of-plane orientation, with weaker (222) reflections also being observed, indicating a highly textured growth with limited secondary orientation.

Rocking curve (RC) measurements, shown in figure 3, were performed to further evaluate the crystal quality. The full width at half maximum (FWHM) of the Pt(111) reflection decreases from 3.0° for the film deposited at 180 W to 0.2° for the film deposited at 90 W, indicating a substantial improvement in crystalline quality and reduced mosaic spread under lower deposition power. A shoulder is evident in the RC measurement of the higher quality film, figure 2(b), which indicates the presence of STO with wider RC FWHM.

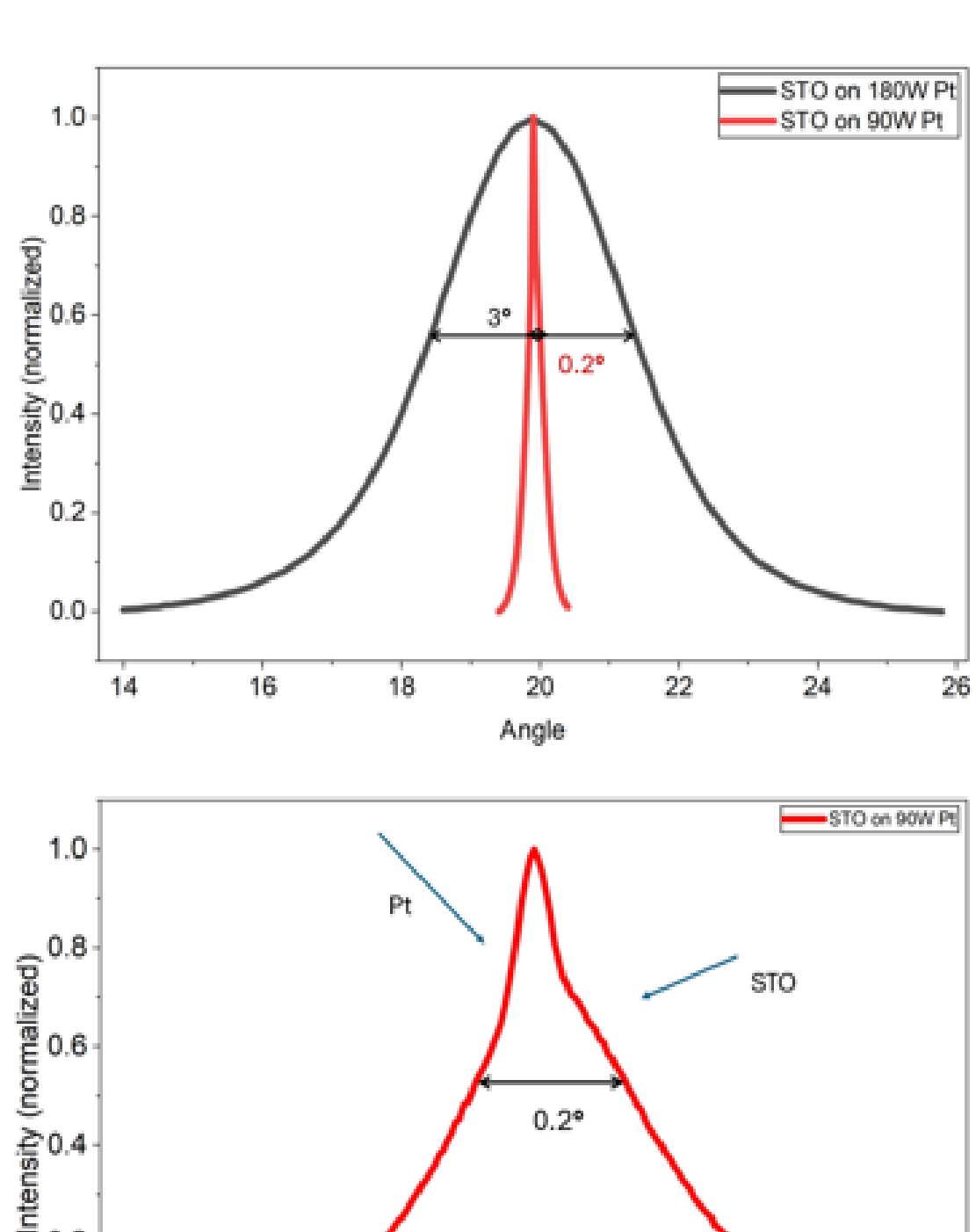

*Figure 3 Rocking Curve for (a) STO on Pt samples grown at 90W (0.27° FWHM) and 180W (3° FWHM) (b) STO sample on Pt grown at 90W.*

The expected diffraction peaks for Pt(111) and STO(111) are located at 2θ values of 39.763° and 39.984°, respectively. Due to the similar crystal structures and closely matched lattice constants of Pt (3.92 Å) and STO (3.905 Å), the corresponding diffraction peaks are not clearly resolved, resulting in an apparent overlap of the Pt and STO reflections. As shown in figure 3, no distinct STO peak is observed separately from the Pt peak, suggesting that the STO film adopts the crystallographic orientation of the underlying Pt layer.

*Surface Structure*

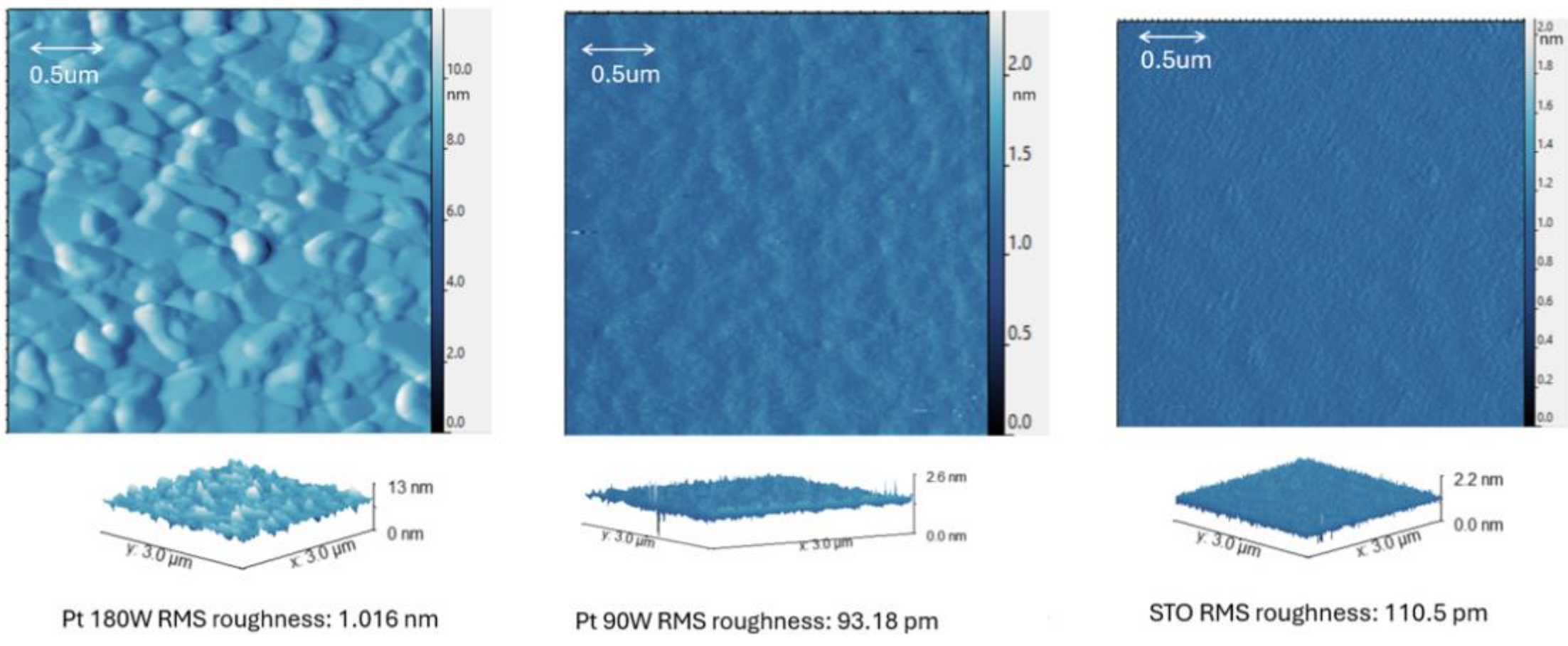

*Figure 4: Atomic force microscopy of, (a) Pt 180 W growth, (b) Pt 90 W growth, (c) 30nm-thick SrTiO3 on Pt/($\mathbf{1\bar{1}02}$) $Al_2O_3$.*

Figure 4 presents AFM topography images of Pt films grown at sputtering powers of 180 W (4a) and 90 W (4b), together with the surface morphology of the STO film (4c) grown on a 90W Pt template. All AFM measurements were performed in contact mode over a scan area of 3 × 3 μm² with a lateral resolution of 512 × 512 pixels. The Pt film deposited at 180 W has an RMS surface roughness of approximately 1 nm, with pronounced surface features and large-scale morphological artifacts evident across the scanned area. In contrast, the Pt film grown at 90 W shows a significantly smoother surface morphology, with no observable microcracks, pinholes, or grain boundaries. A similarly smooth surface morphology is observed for the STO film. The STO surface exhibits an RMS roughness of 110.5 pm, indicative of an atomically smooth film when grown on a Pt (111) substrate with similar roughness (93.2pm). The STO film morphology closely follows that of the underlying Pt surface across the investigated growth conditions, with morphological variations arising only from changes in the Pt surface structure.

This enhancement in surface structure is consistent with the XRD RC results, which show a significantly decreased FWHM for the 90 W Pt film. Together, these results demonstrate that improved Pt crystalline quality and surface smoothness directly promote the high-quality, highly oriented growth of STO films.

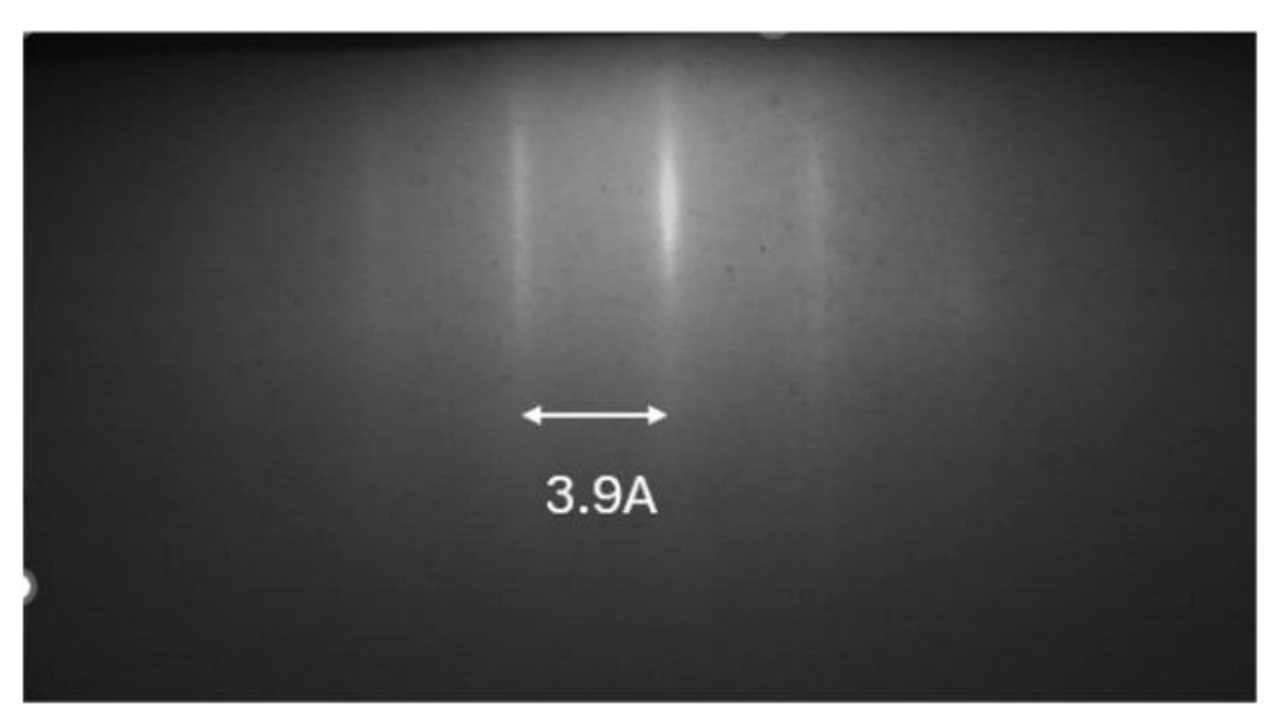

*Figure 5: Reflection high-energy electron diffraction of STO/Pt/$Al_2O_3$ multilayer.*

To further elucidate the crystalline structure of the STO film, reflection high-energy electron diffraction (RHEED) measurements were performed. The RHEED pattern shown in Fig. 6 is consistent with the XRD results and exhibits a streaky, weakly spot-modulated pattern characteristic of STO (111) orientation. The presence of sharp and well-defined streaks indicates a high degree of in-plane ordering over the substrate surface and suggests a quasi–two-dimensional growth mode. $k_{||}$ was found via xRHEED python simulation and used to calculate lattice constant[16].

## Conclusions

Near epitaxial (111)SrTiO (STO) thin films were grown successfully by rf reactive magnetron sputtering on ((111)Pt/ ( )$Al_2O_3$ substrates. The growth temperature and power of Pt growth were found to play an important role in controlling the crystallographic orientations and surface morphology of the STO films on Pt/Al2O3 substrates. The STO film thickness ranged from 30nm to 60nm. X-ray photoelectron spectroscopy verified the chemical composition of the STO films. XRD results reveal that the STO thin films had (111) crystal orientation. XRD rocking curve measurements of the STO films have a narrow full width at half maximum of 0.2°, demonstrating high film quality. The STO films have a surface roughness of 110pm, which is less than the 390pm lattice constant of STO. The STO surface appears extremely smooth, and no spotty features associated with structural defects are present in the RHEED images. With additional investigation, these STO thin films can be used as dielectric material for capacitor structures with high capacitance (8.85-13.28 μF/cm$^2$).

**Author contributions**

Ekaterina A. Taylor conducted the experiments, characterized the performance of the devices and materials, and wrote the original draft. Ethan R. Cronk contributed with XRD characterisation and STO growth and Kalani Perera contributed to the Pt growth. Nuri Emanetoglu and Nicholas S. Bingham contributed resources, project supervision, and contributed to the conceptualization, formal analysis, project administration, and writing – review & editing of the manuscript.

**Acknowledgements**

This material is based upon work supported by the U.S. Department of Energy, Office of Science, Office of Basic Energy Sciences Established Program to Stimulate Competitive Research (EPSCoR) under Award Number DE-SC0021981.